\documentclass[twocolumn,showpacs,preprintnumbers,amsmath,amssymb,aps,letterpaper]{revtex4}

\input{psfig.sty}

\newcommand{\be}{\begin{equation}}
\newcommand{\ee}{\end{equation}}
\newcommand{\bea}{\begin{eqnarray}}
\newcommand{\eea}{\end{eqnarray}}

   \usepackage{color}

\begin{document}
\addtolength{\topmargin}{1.3cm}
\title{Statefinder diagnostic for the modified polytropic Cardassian universe}
\author{Ze-Long Yi}
\author{Tong-Jie Zhang}
\email{tjzhang@bnu.edu.cn}

\address{Department of Astronomy, Beijing Normal University,
Beijing, 100875, P.R.China}

\begin{abstract}
We apply the statefinder diagnostic to the modified polytropic
Cardassian universe in this work. We find that the statefinder
diagnostic is quite effective to distinguish Cardassian models from
a series of other cosmological models. The $s-r$ plane is used to
classify the modified polytropic Cardassian models into six cases.
The evolutionary trajectories in the $s-r$ plane for the cases with
different $n$ and $\beta$ reveal different evolutionary properties
of the universe. In addition, we combine the observational $H(z)$
data, the cosmic microwave background data and the baryonic acoustic
oscillation data to make a joint analysis. We find that \textbf{Case
2} can be excluded at the 68.3\% confidence level and any case is
consistent with the observations at the 95.4\% confidence level.

\end{abstract}

\pacs{98.80.Es,95.35.+d,98.80.Jk}

\maketitle

\section{Introduction}
\label{sec:intro}

Recent type Ia Supernova observations, along with other observations
such as cosmic microwave background (CMB) and galaxy power spectra,
support the fact that the present expansion of our universe is
accelerating \cite{Riess,Perlmutter}. In order to explain the
accelerated expansion of the universe, cosmologists have tried to
explore many cosmological models. A negative pressure term called
dark energy is always taken into account, such as the cosmological
constant model with equation of state $\omega_{\rm DE}=p_{\rm
DE}/\rho_{\rm DE}=-1$ where $p_{\rm DE}$ and $\rho_{\rm DE}$ are
pressure and density of the dark energy, respectively
\cite{Carroll}, the quiessence whose equation of state $\omega_{\rm
Q}$ is a constant between -1 and -1/3 \cite{Alam}, and the
quintessence which is described in terms of a scalar field $\phi$
\cite{Caldwell,Sahni1}. Some other candidates are constructed in
different ways, such as the braneworld models which explain the
acceleration through the fact that the general relativity is
formulated in 5 dimensions instead of the usual 4 \cite{Csaki}, and
the Cardassian models which investigate the acceleration of the
universe by a modification to the Friedmann-Robertson-Walker (FRW)
equation \cite{Freese}. In these cases, the dark energy component is
not involved, but the accelerated expansion can still be obtained.
The quantity $\omega_{\rm eff}$ is usually described as an effective
equation of state for these models, and can be expressed by the
Hubble parameter $H$ and its derivatives with respect to redshift
$z$ \cite{Alam}.

As so many cosmological models have been developed, a discrimination
between these contenders becomes necessary. In order to achieve this
aim, Sahni et al. proposed a new geometrical diagnostic named the
statefinder pair $\{r, s\}$, where $r$ is generated from the scalar
factor $a$ and its derivatives with respect to the cosmic time $t$,
just as the Hubble parameter $H$ and the deceleration parameter $q$,
and $s$ is a simple combination of $r$ and $q$ \cite{Sahni2}. The
statefinder pair has been used to discriminate a series of
cosmological models, including the LCDM universe with a cosmological
constant $\Lambda$ and a cold dark matter term (CDM), the Chaplygin
gas, the holographic dark energy models, the quintessence, the
braneworld models and so on. Clear differences for the evolutionary
trajectories in the $s-r$ plane have been found.

In this paper, we apply the statefinder diagnostic to the modified
polytropic Cardassian universe. The original Cardassian model based
on two parameters $\Omega_{m0}$ and $n$ was first suggested by
Freese \& Lewis \cite{Freese}. It was generated from a modification
to the Friedmann equation. Such a universe is spatially flat and
accelerating today. But it involves no dark energy term and is
dominated by merely matter and radiation. These Cardassian models
predict the same distance-redshift relation as generic quintessence
models, although they generate from completely different physical
principles. A generalized Cardassian model--the modified polytropic
Cardassian universe can be obtained by introducing an additional
parameter $\beta$ into this model \cite{Gondolo1,Gondolo2,Wang},
which reduces to the original model if $\beta=1$. The
distance-redshift relation predictions of generalized Cardassian
models can be very different from generic quintessence models, and
can be differentiated with data from surveys of Type Ia Supernovae
such as SuperNova/Acceleration Probe (SNAP).
In all, the modified polytropic Cardassian universe can predict more
fresh physical information than the original Cardassian. It is
worthy of more detailed discussions.

In this work, we successfully classify the modified polytropic
Cardassian models into six cases by the statefinder diagnostic. The
cases with different $n$ and $\beta$ correspond to different
evolutionary trajectories in the $s-r$ plane. The fact that LCDM
corresponds to a fixed point (0, 1) in the $s-r$ plane plays a
significant role for our classification. Also, it is very important
to find where the evolutionary trajectories start and end. Another
key standpoint is whether the evolutionary trajectory has a crossing
with LCDM. We also study the relation between $(n, \beta)$ and the
crossing redshift $z_{\rm C}$, at which the modified polytropic
Cardassian universe intersects with LCDM in the $s-r$ plane. We find
that the modified polytropic Cardassian models can be distinguished
from other independent cosmological models by the statefinder
diagnostic.

In addition, we use the observational $H(z)$ data derived from ages
of the passively evolving galaxies \cite{Simon}, the newly measured
value of the CMB shift parameter $\mathcal {R}$ \cite{Odman} and the
$\mathcal {A}$-parameter which describes the baryonic acoustic
oscillation (BAO) peak \cite{Eisenstein} to make a combinational
constraint. We assume a prior of $H_0=72\pm8$ suggested by the
Hubble Space Telescope (HST) Key Project \cite{Freedmann}. From the
confidence regions, we find that \textbf{Case 2} is not consistent
with the observations at the 68.3\% confidence level and all the six
cases do not conflict with the observations at the 95.4\% confidence
level.

This paper is organized as follows: In Sec.2, we briefly review the
modified polytropic Cardassian universe. In Sec.3, we introduce the
statefinder pair $\{r, s\}$. In Sec.4, we apply the statefinder
diagnostic to various Cardassian models. In Sec.5, we make a
combinational constraint on the parameters of the modified
polytropic Cardassian universe. In Sec.6, the discussions and the
conclusions are given.

\section{The Modified Polytropic Cardassian Universe}
Measurements of CMB suggest a flat geometry for our universe
\cite{Halverson,Netterfield}. If we consider a spatially flat FRW
universe, the basic equation can be written as
\begin{equation}
H^2=\frac{8\pi G}{3}\rho, \label{eq1}
\end{equation}
where $G$ is Newton's universal gravitation constant and $\rho$ is
the density of summation of both matter and vacuum energy. Freese \&
Lewis \cite{Freese} proposed a model called the Cardassian universe
by adding a term on the right side of Eq.(\ref{eq1}),
\begin{equation}
H^2=\frac{8\pi G}{3}\rho_{\rm m}+B\rho_{\rm m} ^n,\label{eq2}
\end{equation}
where $n$ is assumed to satisfy $n<2/3$ and $\rho_{\rm m}$ is
always taken as a contribution of only the matter (in this paper
we do not plan to consider radiation). If $n=0$, it is identical
to the cosmological constant universe.
If $B=0$, it reduces to the usual FRW equation, but with density of
only matter. Thus, it is easy to get a new expression of $H$ from
Eq.(\ref{eq2}),
\begin{equation}
H^2=H_{\rm 0}^2[\Omega_{m0}(1+z)^3+(1-\Omega
_{m0})(1+z)^{3n}],\label{eq3}
\end{equation}
by using
\begin{equation}
\rho_{\rm m}=\rho_{\rm m0}(1+z)^3=\Omega_{m0}\rho_{\rm
c}(1+z)^3,\label{eq4}
\end{equation}
where $\rho_{\rm m0}$ is current value of $\rho_{\rm m}$ and
$\rho_{\rm c}=3H_{\rm 0}^2/8\pi G$ is the critical density of the
universe. Clearly, this model predict the same distance-redshift
relation as the quiessence with $\omega_{\rm Q}=n-1$. But we notice
that they have completely different essentials. The quiessence
requires a dark energy component while the Cardassian does not.

In the earlier time, the universe is dominated by the first term in
Eq.(\ref{eq3}) and the ordinary FRW behavior works. The additional
term which consists of only matter gradually becomes a dominant
driver afterwards. This transition was found to occur at $z\sim
\emph{O}(1)$ \cite{Freese}. Then the universe is caused to
accelerate. The period of acceleration for this model is usually
called the Cardassian era. The Cardassian model is attractive
because the universe is flat and accelerating today but no vacuum
energy is involved. And it has been demonstrated compatible with a
series of observational tests, including the CMB data, the age of
the universe, the structure formation and the cluster baryon
fraction \cite{Freese}.

The modified polytropic Cardassian universe can be obtained by
introducing an additional parameter $\beta$ into the above model
\cite{Gondolo1,Gondolo2,Wang},
\begin{equation}
H^2=H_{\rm 0}^2[\Omega_{m\rm 0}(1+z)^3+(1-\Omega_{m\rm 0})f_{\rm
X}(z)],\label{eq5}
\end{equation}
where
\begin{equation}
f_{\rm X}(z)=\frac{\Omega_{m\rm 0}}{1-\Omega_{m\rm
0}}(1+z)^3[(1+\frac{\Omega_{m\rm
0}^{-\beta}-1}{(1+z)^{3(1-n)\beta}})^{1/\beta}-1].\label{eq6}
\end{equation}
The two parameters $(n, \beta)$ are
usually taken as $n<2/3$ and $\beta>0$. If $\beta=1$, the model
reduces to the original one characterized by Eq.(\ref{eq3}) while if
$f_{\rm X}(z)=1$, the model just corresponds to LCDM.

Similar to Wang et al., we also take $\Omega_{m\rm 0}=0.3$ as a
prior in subsequent discussions.
In the work of Wang et al., this model was compared with current
supernova data and CMB data. It was proved that the existing data
can be well fit for several chosen values of $n$ and $\beta$. Also,
the simulated data were constructed to make a discrimination between
the modified polytropic Cardassian universe and LCDM as well as the
quintessence. Once $\Omega_{m\rm 0}$ is known with an accuracy of
10\%, SNAP can determine the sign of the time dependence of dark
energy density which provides a first discrimination between various
cosmological models \cite{Wang}. In this work, we use a geometrical
tool-the statefinder diagnostic to make a classification and a
discrimination about the modified polytropic Cardassian models.

\section{A Brief Overview of the Statefinder Diagnostic}
The Hubble parameter $H=\dot{a}/a$ and the deceleration parameter
$q=-\ddot{a}/aH^2$ are two traditional geometrical diagnostics. They
only depend on the scalar factor $a$ and its derivatives with
respect to $t$, i.e., $\dot{a}$ and $\ddot{a}$. Through $a=1/(1+z)$,
the deceleration parameter $q$ can be expressed as
\begin{equation}
q(z)=\frac{H'}{H}(1+z)-1,\label{eq7}
\end{equation}
where $H'$ is the derivative of $H$ with respect to redshift $z$.
The deceleration parameter $q$ is a good choice to describe the
expansion state of our universe, but it is not perfect enough to
characterize the cosmological models uniquely. This shortage can be
easily seen from the fact that many models may correspond to the
same current value of $q$.
And this difficulty can be overcome by another geometrical
diagnostic--the statefinder pair $\{r, s\}$. This approach has been
used to distinguish a series of cosmological models successfully.

For a spatially flat universe, the statefinder $r$ is defined as
follows \cite{Sahni2}
\begin{equation}
r=\frac{\stackrel{\cdots}{a}}{aH^3},\label{eq8}
\end{equation}
where $\stackrel{\cdots}{a}$ is the third derivative of $a$ with
respect to $t$. $s$ is just a combination of $r$ and $q$,
\begin{equation}
s=\frac{r-1}{3(q-1/2)}.\label{eq9}
\end{equation}
The statefinder pair was first introduced to analyze a flat universe
with a cold matter and a dark energy term. For these contenders, $r$
is given by
\begin{equation}
r=1+\frac{9}{2}\omega_{\rm DE}(1+\omega_{\rm DE})\Omega_{\rm
DE}-\frac{3}{2}\frac{\dot{\omega}_{\rm DE}}{H}\Omega_{\rm
DE}.\label{eq10}
\end{equation}
where $\Omega_{\rm DE}=\rho_{\rm DE}/\rho_{\rm c}$
and $\dot{\omega}_{\rm DE}$ is the derivative of $\omega_{\rm DE}$
with respect to $t$. And the other diagnostic,
\begin{equation}
s=1+\omega_{\rm DE}-\frac{1}{3}\frac{\dot{\omega}_{\rm
DE}}{\omega_{\rm DE}H}.\label{eq11}
\end{equation}
From the two equations above it is easy to realize that LCDM
corresponds to a fixed point (0, 1) in the $s-r$ plane and the
standard cold dark matter (SCDM) universe with no dark energy term
locates at (1, 1) forever. For this particularity, the current
values of $\{r, s\}$ provide a considerable way to measure the
distance from a specific model to LCDM.

More generally, $r$ and $s$ can be given in terms of the Hubble
parameter $H$ and its first and second derivatives $H'$ and $H''$
with respect to redshift $z$,
\begin{equation}
r(z)=1-2\frac{H'}{H}(1+z)+{\frac{H''}{H}+(\frac{H'}{H})^2},\label{eq12}
\end{equation}
\begin{equation}
s(z)=\frac{-2H'(1+z)/H+{H''/H+(H'/H)^2}}{3(H^{'}(1+z)/H-3/2)}.\label{eq13}
\end{equation}
Thus we can use the new tool to describe the evolutionary
trajectories of the modified polytropic Cardassian universe.

\section{The $s-r$ Plane for Modified Polytropic Cardassian Universe}
For the original Cardassian model with $\beta=1$, it corresponds to
a compatible expression with the quiessence with $\omega_{\rm
Q}=n-1$. The quiessence has been studied using the statefinder
diagnostic in several literatures \cite{Alam,Sahni2}. In the $s-r$
plane, a vertical line with $s=n$ and $r$ changing monotonically
from 1 to $1+9n(n-1)/2$ represents the evolutionary trajectory of
the universe.

For cases with $\beta\neq1$, we first pay our attention to the
epoches in the far past $(a\to0)$ and the far future
$(a\to+\infty)$. It is clear that $z\to+\infty$ represents the
former case and $z\to-1$ stands for the latter.
From Eq.(\ref{eq12}), it is easy to find the limit condition
\begin{equation}
\lim_{z\to+\infty}r(z)=1, \label{eq14}
\end{equation}
which means that the value of $r$ in the far past is independent on
$n$ and $\beta$. However, we can not derive the similar properties
for the value of $s$ from Eq.(\ref{eq13}). It is related to both of
$n$ and $\beta$. For the limit of $z\to-1$, from Eq.(\ref{eq12}), we
get
\begin{equation}
\lim_{z\to -1}r(z)=1+\frac{9}{2}n(n-1). \label{eq15}
\end{equation}
And from Eq.(\ref{eq13}), we have
\begin{equation}
\lim_{z\to -1}s(z)=n. \label{eq16}
\end{equation}
The values of both $s$ and $r$ are independent on $\beta$.

Generally,
some models have a crossing with LCDM in the $s-r$ plane.
By substituting the Hubble parameter, Eq.(\ref{eq5}), in the
expressions of $r$ and $s$, Eq.(\ref{eq12}) and Eq.(\ref{eq13}), we
find that the crossing with LCDM happens at the redshift
\begin{equation}
z_{\rm C}=[(\Omega_{\rm
m0}^{-\beta}-1)\frac{-n}{1-\beta+n\beta}]^{\frac{1}{3(1-n)\beta}}-1.\label{eq17}
\end{equation}
And we notice that the crossing can happen only if the following
inequality is satisfied,
\begin{equation}
(\Omega_{\rm m0}^{-\beta}-1)\frac{-n}{1-\beta+n\beta}\geq0.
\label{eq18}
\end{equation}
Due to the prior $\Omega_{\rm m0}=0.3$ and $\beta>0$, $\Omega_{\rm
m0}^{-\beta}-1>0$ is naturally satisfied. Thus we may describe the
condition in Eq.(\ref{eq18}) equivalently as
\begin{equation}
\frac{n}{1-\beta+n\beta}\leq0. \label{eq19}
\end{equation}

\begin{figure}[!t]
\centerline{\psfig{file=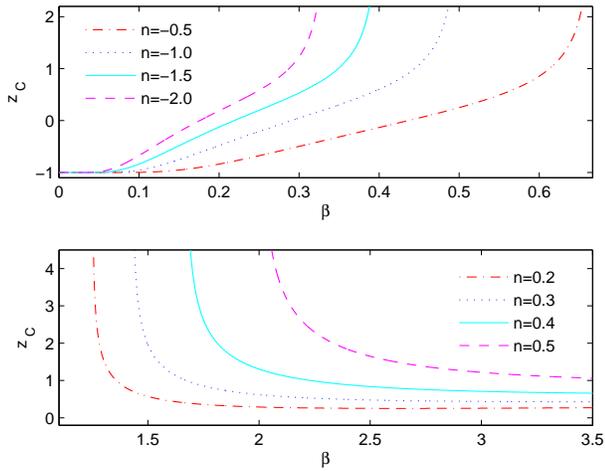,width=3.6in,angle=0}} \caption{
$\beta-z_{\rm C}$ planes for $n$=-0.5, -1.0, -1.5 and -2.0 (the top
panel) as well as $n$=0.2, 0.3, 0.4 and 0.5 (the bottom
panel).}\label{fig1}
\end{figure}

In order to understand the relation among $n$, $\beta$ and $z_{\rm
C}$ clearly, we draw a $\beta-z_{\rm C}$ plane for several fixed $n$
in Fig.\ref{fig1}. For $n>0$, the crossing takes place at $z_{\rm
C}>0$ for any $\beta$, and $z_{\rm C}$ changes little for larger
values of $\beta$. For $n<0$, whether $z_{\rm C}>0$ or $z_{\rm C}<0$
depends on the values of $\beta$. And $z_{\rm C}$ is nearly equal to
-1 if $\beta$ is small enough. For the particular case which
satisfies $1-\beta+n\beta=0$, the expected crossing happens at
$z_{\rm C}\to\infty$. Clearly, $n=0$ and $\beta=1/(1-n)$ are the two
critical conditions.


\begin{figure}[!t]
\centerline{\psfig{file=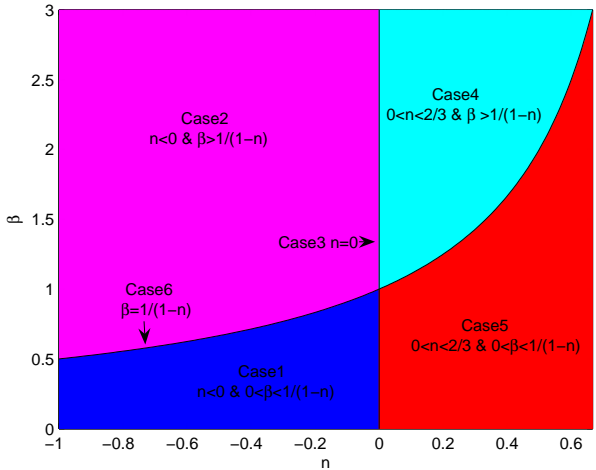,width=3.6in,angle=0}} \caption{
$n-\beta$ plane based on $n$ and $\beta$. The vertical solid line
represents $n=0$ (\textbf{Case 3}), and the declined solid curve
stands for $\beta=1/(1-n)$ (\textbf{Case 6}).}\label{fig2}
\end{figure}

Now we use Eq.(\ref{eq19}) to classify the modified polytropic
Cardassian universe. We draw an $n-\beta$ plane in Fig.\ref{fig2}
and divide it into different regions by the critical curves $n=0$
and $\beta=1/(1-n)$:
\begin{equation}
\left\{
\begin{array}{llllll}
{\textbf{Case 1}} & n<0\ {\rm and}\ 0<\beta<1/(1-n);\\
{\textbf{Case 2}} & n<0\ {\rm and}\ \beta>1/(1-n);\\
{\textbf{Case 3}} & n=0;\\
{\textbf{Case 4}} & 0<n<2/3\ {\rm and}\ \beta>1/(1-n);\\
{\textbf{Case 5}} & 0<n<2/3\ {\rm and}\ 0<\beta<1/(1-n);\\
{\textbf{Case 6}} & \beta=1/(1-n).
\end{array}\right.\label{density}
\end{equation}

\begin{figure}[!t]
\centerline{\psfig{file=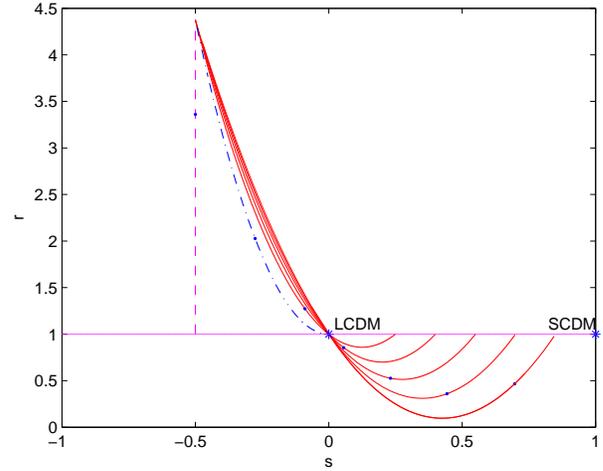,width=3.6in,angle=0}}
\caption{$s-r$ plane for \textbf{Case 1}. The star (0, 1)
corresponds to LCDM, the star (1, 1) SCDM and other dots the current
values of the Statefinder pair, i.e., $(s_{\rm 0}, r_{\rm 0})$ (also
for subsequent cases). And the solid curves from bottom to top
correspond to the evolutionary trajectories of $n$=-0.5,
$\beta$=0.1, 0.2, 0.3, 0.4 and 0.5 respectively.
The dot-dash curves represent the critical case $\beta=2/3$. For a
comparison, the vertical dashed line stands for the case of
$\beta=1$.}\label{fig3}
\end{figure}

\textbf{Case 1} ($n<0$ and $0<\beta<1/(1-n)$): $n$=-0.5,
$\beta$=0.1, 0.2, 0.3, 0.4 and 0.5 are taken for a qualitative
analysis. We plot the $s-r$ plane in Fig.\ref{fig3}. Although the
original Cardassian model with $\beta=1$ is not involved in this
case, we still draw this curve for a comparison. All the curves
start at $r=1$ and $0<s<1$ on the horizontal line, i.e., on the
right of LCDM. After passing by LCDM, they arrive at their common
end $(-0.5, 4.38)$ in far future. And the crossing with LCDM happens
at $z_{\rm C}$=-0.999, -0.84, -0.50, -0.13 and 0.25 for $\beta$=0.1,
0.2, 0.3, 0.4 and 0.5 respectively. Particularly, the fixed point
(0, 1) for LCDM just is the beginning point for the critical case of
$\beta=1/(1-n)=2/3$.


\begin{figure}[!t]
\centerline{\psfig{file=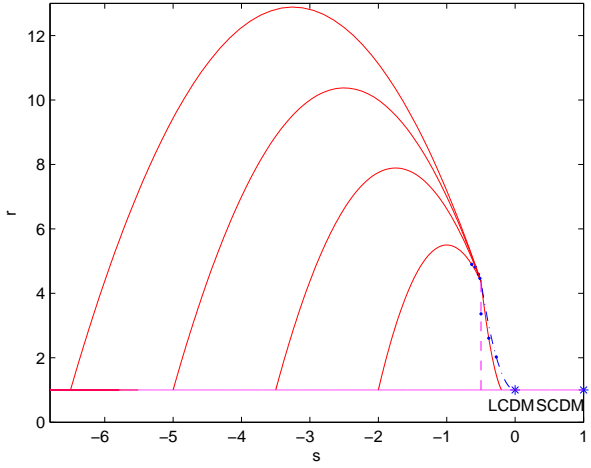,width=3.6in,angle=0}}
\caption{$s-r$ plane for \textbf{Case 2}. The solid curves from
right to left correspond to the evolutionary trajectories of
$n$=-0.5, $\beta$=0.8, 2, 3, 4 and 5 respectively. And the vertical
dashed line stands for the case of $\beta=1$.
Same as Fig.\ref{fig3}, the dot-dash curve represents the critical
case $\beta=2/3$.}\label{fig4}
\end{figure}

\textbf{Case 2} ($n<0$ and $\beta>1/(1-n)$): To be consistent with
\textbf{Case 1}, we consider $n=-0.5$ again, but $\beta$=0.8, 1, 2,
3, 4 and 5 respectively. The $s-r$ plane is plotted in
Fig.\ref{fig4}.
The crossing with LCDM never occurs for this case. The vertical line
$s=n$ for the evolutionary trajectory of $\beta=1$ divides the whole
plane into two parts. The left part with $n<s<0$ stands for
$1/(1-n)<\beta<1$ and the right part with $s<n$ corresponds to
$\beta>1$. And all the evolutionary trajectories commence at one
point with $r=1$. The same as \textbf{Case 1}, they arrive at
$(-0.5, 4.38)$ in the end.


\begin{figure}[!t]
\centerline{\psfig{file=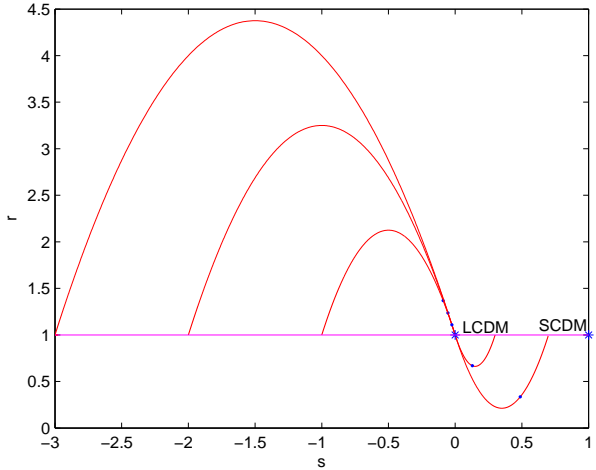,width=3.6in,angle=0}}
\caption{$s-r$ plane for \textbf{Case 3}. The solid curves from
bottom to top stand for $\beta$=0.3, 0.7, 2, 3 and 4 respectively.
The case of $\beta=1$ just corresponds to the fixed point for LCDM,
i.e., (0, 1).}\label{fig5}
\end{figure}

\textbf{Case 3} ($n=0$): We plot the $s-r$ plane for $\beta$=0.3,
0.7, 1, 2, 3 and 4 in Fig.\ref{fig5} . One point with $r=1$ acts as
the beginning point, and (0, 1) the common end. This diagram is
divided into two segments by the LCDM fixed point, or equivalently
the modified polytropic Cardassian model with $n=0$ and $\beta=1$.
The segment with $r>1$ corresponds to $\beta>1$ while the other
segment with $r<1$ corresponds to $\beta<1$.


\begin{figure}[!t]
\centerline{\psfig{file=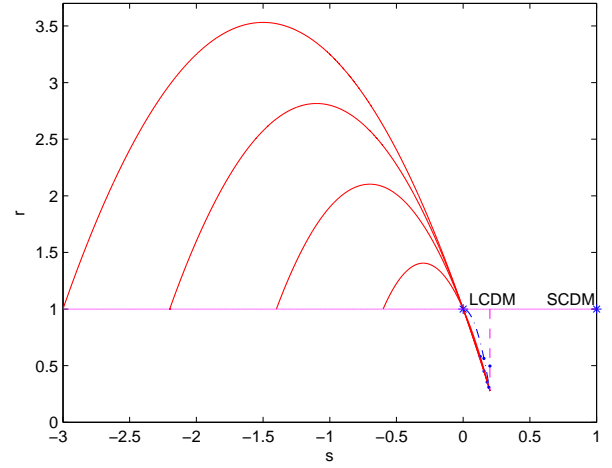,width=3.6in,angle=0}}
\caption{$s-r$ plane for \textbf{Case 4}. The solid curves from
right to left stand for the evolutionary trajectories of $n=0.2$,
$\beta$=2, 3, 4 and 5 respectively.
The dot-dash curve represents the critical case $\beta=5/4$. For a
comparison, the vertical dashed line stands for the case of
$\beta=1$.}\label{fig6}
\end{figure}

\textbf{Case 4} ($0<n<2/3$ and $\beta>1/(1-n)$): The $s-r$ plane for
$n=0.2$, $\beta$=2, 3, 4 and 5 is plotted in Fig.\ref{fig6}.
Clearly, all the curves
start with $r=1$
and $s<0$, and then pass by LCDM after an arc route. All the latter
evolutionary parts of the trajectories nearly overlap with each
other. They are insensitive to $\beta$. The point $(0.2, 0.28)$ is
the common end. The crossing redshifts are not too far from each
other for different values of $\beta$, i.e., $z_{\rm C}\simeq 0.3$.
It is interesting that the fixed point (0, 1) for LCDM is just the
beginning point for the critical case of $\beta=1/(1-n)=5/4$. And
the Cardassian models for this case can satisfy the weak energy
condition $\omega=p_{\rm X}/(\rho_{\rm m}+\rho_{\rm X})>-1$ although
the effective equation of state satisfies $\omega_{\rm eff}=p_{\rm
X}/\rho_{\rm X}<-1$. $w_{\rm eff}<-1$ is consistent with many
observations such as CMB and large scale structure data
\citep{Schuecker,Melchiorri}.



\begin{figure}[!t]
\centerline{\psfig{file=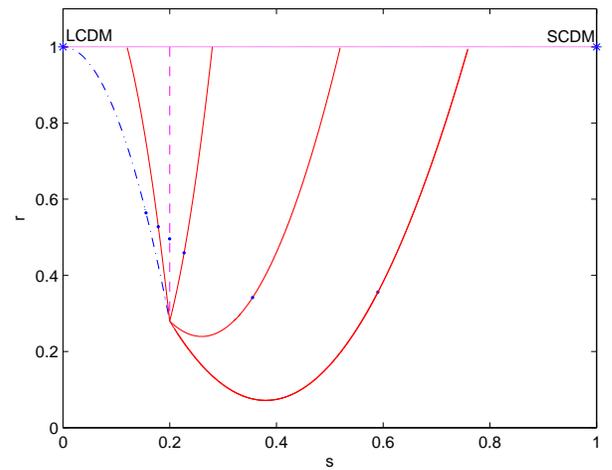,width=3.6in,angle=0}}
\caption{$s-r$ plane for \textbf{Case 5}. The solid curves from
right to left correspond to the evolutionary trajectories of
$n$=0.2, $\beta$=0.3, 0.6, 0.9 and 1.1 respectively. And the
vertical dashed line stands for the case of $\beta=1$.
Same as Fig.\ref{fig6}, the dot-dash curve represents the critical
case $\beta=5/4$.}\label{fig7}
\end{figure}

\textbf{Case 5} ($0<n<2/3$ and $0<\beta<1/(1-n)$): Same as
\textbf{Case 4}, $n=0.2$ is considered, but $\beta$=0.3, 0.6, 0.9,
1.0 and 1.1. We plot the $s-r$ plane in Fig.\ref{fig7}.
All the evolutionary trajectories start with $r=1$ and $s>0$. They
arrive at $(0.2, 0.28)$ in the end, and never pass by LCDM. The
whole plane is divided into two parts by the vertical line $s=n$
(corresponding to the case of $\beta=1$). The left part satisfies
$1<\beta<5/4$ and $\beta<1$ is satisfied for the right.


\begin{figure}[!t]
\centerline{\psfig{file=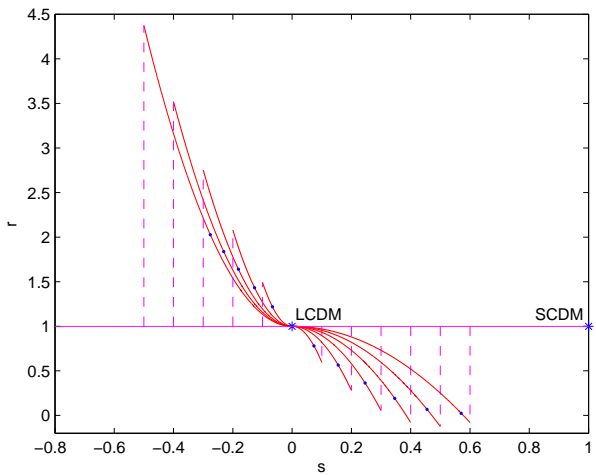,width=3.6in,angle=0}}
\caption{$s-r$ plane for \textbf{Case 6}. The point (0, 1) stands
for $n=0$. Then $n$ increases by a interval of 0.1 until 0.6 for
cases on the right of (0, 1) and $n$ decreases by a interval of 0.1
until -0.5 for cases on the left of (0, 1). The other vertical
dashed lines stand for the evolutionary trajectories of cases with
$\beta=1$ and different values of $n$.}\label{fig8}
\end{figure}

\textbf{Case 6} ($\beta=1/(1-n)$): This is another critical case
besides \textbf{Case 3}. $n$=-0.5, -0.4, -0.3, -0.2, -0.1, 0, 0.1,
0.2, 0.3, 0.4, 0.5, and 0.6 are considered. For this case, crossing
with LCDM happens at $z_{\rm C}\to\infty$. We plot the $s-r$ plane
in Fig.\ref{fig8}. All the evolutionary trajectories start at LCDM
in the past and end at
$(-1+3n/2, 1+9n(n-1)/2)$ in the future.


We have successfully given a qualitative analysis to the modified
polytropic Cardassian universe from the statefinder viewpoint and it
has been classified into six cases. A prior $\Omega_{\rm m0}=0.3$ is
adopted in this work. We also take some other values of $\Omega_{\rm
m0}$ for comparison and notice that $\Omega_{\rm m0}$ is not a
sensitive parameter for our analysis in this work. And it has been
clear that the $s-r$ plane is effective enough to discriminate the
modified polytropic Cardassian models. The fact that LCDM
corresponds to a fixed point (0, 1) in the $s-r$ plane plays a
significant role for our classification. Whether the evolutionary
trajectory has a crossing with LCDM is our basic standpoint.
Also, the beginning points and the ending points are both key
factors for shapes of the evolutionary trajectories.


\section{Data Analysis from Observational $H(z)$
Data, CMB and BAO}
In order to understand the six cases above more clearly, we use the
observational $H(z)$ data derived from the passively evolving
galaxies, the cosmic microwave background (CMB) shift parameter
$\mathcal {R}$ and the $\mathcal {A}$-parameter which describes the
baryonic acoustic oscillation (BAO) peak to make a joint analysis.

The Hubble parameter $H(z)$ depends on the differential age of the
universe in this form
\begin{equation}
H(z)=-\frac{1}{1+z}\frac{{\rm d}z}{{\rm d}t},\label{eq21}
\end{equation}
which provides a direct measurement for $H(z)$ through a
determination of ${\rm d}z/{\rm d}t$. By using the differential ages
of passively evolving galaxies determined from the Gemini Deep Deep
Survey \cite{Abraham} and archival data \cite{Treu, Treu+, Nolan,
Nolan+}, Simon et al. determined a set of observational $H(z)$ data
in the range $0\lesssim z\lesssim 1.8$ and used them to constrain
the dark energy potential and its redshift dependence \cite{Simon}.
Yi \& Zhang first first used them to analyze the holographic dark
energy models and got a consistent result with others \cite{Yi}.


The model-independent shift parameter $\mathcal {R}$ can be derived
from CMB data. It is defined as \cite{Odman}
\begin{equation}
\mathcal {R}=\sqrt{\Omega_{\rm m0}}\int_0^{z_{\rm
r}}\frac{dz}{E(z)},\label{eq22}
\end{equation}
where $E(z)=H(z)/H_0$ and $z_{\rm r}=1089$ is the redshift of
recombination. From the three-year result of WMAP \cite{Spergel},
Wang \& Mukherjee  estimated $\mathcal {R}=1.70\pm0.03$
\cite{Wang+}.

Using a large spectroscopic sample of 46748 luminous red galaxies
covering 3816 square degrees out to $z=0.47$ from the Sloan Digital
Sky Survey, Eisenstein et al. \cite{Eisenstein} successfully found
the acoustic peaks in the CMB anisotropy power spectrum, described
by the model-independent $\mathcal {A}$-parameter,
\begin{equation}
\mathcal {A}=\sqrt{\Omega_{\rm m0}}
[\frac{1}{z_1E^{1/2}(z_1)}\int_0^{z_1}\frac{dz}{E(z)}
]^{2/3},\label{eq11}
\end{equation}
where $z_1=0.35$ is the redshift at which the acoustic scale has
been measured. Eisenstein et al. \cite{Eisenstein} suggested the
measured value of the $\mathcal {A}$-parameter as $\mathcal
{A}=0.469\pm0.017$.

\begin{figure}[!t]
\centerline{\psfig{file=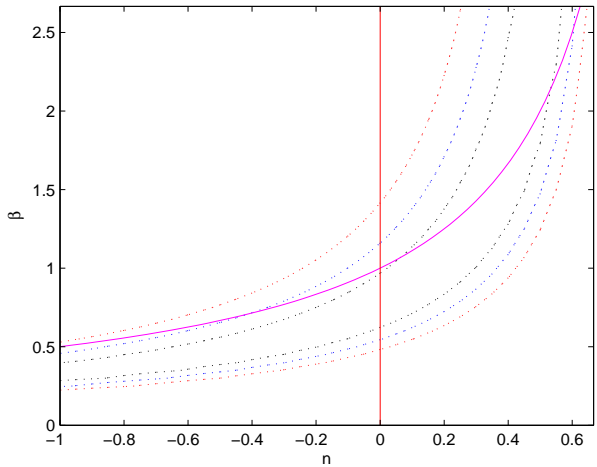,width=3.6in,angle=0}}
\caption{Confidence regions in the $n-\beta$ plane for the joint
analysis (the regions from inner to outer stand for confidence
levels at 68.3\%, 95.4\% and 99.7\% respectively). The vertical
solid line represents $n=0$ (\textbf{Case 3}), and the declined
solid curve stands for $\beta=1/(1-n)$ (\textbf{Case
6}).}\label{fig9}
\end{figure}

The best-fit parameters of the modified polytropic Cardassian
universe can be determined by minimizing
\begin{equation}
\chi^2=\sum_i\frac{[H_{\rm th}(z_{i})-H_{\rm ob}(z_{i})]^2}{\sigma
_i^2}+\frac{(\mathcal {R}-1.70)^2}{0.03^2}+\frac{(\mathcal
{A}-0.469)^2}{0.017^2},\label{eq23}
\end{equation}
where $H_{\rm th}(z_i)$ is the theoretical Hubble parameter at
$z_i$, $H_{\rm ob}(z_i)$ is the observational Hubble parameter at
$z_i$ and $\sigma_i$ is the corresponding $1\sigma$ error. We get
the best-fit values $n=-1.85$ and $\beta=0.23$, with $\chi^2_{\rm
min}=10.12$. The best-fit results correspond to \textbf{Case 1}. And
the crossing with LCDM in the $s-r$ plane occurs at $z_{\rm
C}$=0.32. The current values of the diagnostics are $s_{\rm
0}$=-0.19 and $r_{\rm 0}$=1.59. The confidence regions in the
$n-\beta$ plane are plotted in Fig.\ref{fig9}, from which we find
that \textbf{Case 2} can be excluded at the 68.3\% confidence level
and all the six cases are consistent with the observational data at
the 95.4\% confidence level.

\section{Discussions and Conclusions}
The statefinder diagnostic is powerful to discriminate various
cosmological models. Differences of the evolutionary trajectories in
the $s-r$ plane among a series of cosmological models have been
found. For example, LCDM corresponds to a fixed point (0, 1) and the
point (1, 1) represents SCDM.
For the holographic dark energy model with $c=1$ \cite{Cohen}, the
curve in the $s-r$ plane commences at (2/3, 1) in the past and ends
at LCDM in the future, with $s$ monotonically decreasing from 2/3 to
0 and $r$ first decreasing from 1 to a minimum value and then rising
to 1 \cite{Zhang1}.
Both the quintessence tracker models (with tracker potentials
$V=V_{\rm 0}/\phi^\alpha$) and the Chaplygin gas models have arc
evolutionary trajectories, but in different regions
\cite{Alam,Evans,Gorini}. The conditions $-1\leq s\leq0$
and $r\leq1$ are satisfied for the former cosmological model while
$0\leq s\leq1$
and $r\geq1$ for the latter. The evolutionary trajectories of the
coupled quintessence models have more complicated evolutionary
properties \cite{Zhang2}. And quintessence models with other
potentials were studied by Evans et al.\cite{Evans}, also along with
a generalization of $\{r, s\}$ to a non-flat universe.

Also, the $q-r$ plane has been widely used for discussion on the
evolutionary property of the universe. For example, the point (0.5,
1) corresponds to SCDM in the $q-r$ plane and the horizontal
straight line from (0.5, 1) to (-1, 1) stands for LCDM. The
braneworld models have been studied too, including Disappearing Dark
Energy (DDE) as the simplest case \cite{Alam}. LCDM separates the
first braneworld model (named BRANE1 in \cite{Alam} and the
effective equation of state satisfies $\omega_{\rm eff}\leq-1$) from
the second braneworld model (named BRANE2 in \cite{Alam} and the
effective equation of state satisfies $\omega_{\rm eff}\geq-1$) and
DDE models. DDE both begins and ends at SCDM, forming a loop.
However, although both BRANE1 and BRANE2 commence their evolutions
at (0.5, 1) and end at (-1, 1), the diagnostic $r$ satisfies
$r\geq1$ for BRANE1 while $r\leq1$ for BRANE2.

As the $s-r$ plane has been found robust for our classification for
the modified polytropic Cardassian models, we do not intend to make
use of the $q-r$ plane. For the modified polytropic Cardassian
universe, the evolutionary trajectories can be picked out easily
with help of the $s-r$ plane if $\beta\neq1$. In fact, the original
Cardassian model with $\beta$=1 can not be discarded from the
quiessence with $\omega_{\rm Q}=n-1$ because they correspond to
identical expressions for both $r$ and $s$.
Such a consistence has also been mentioned by
Freese \& Lewis, as well as how to distinguish the two models
\cite{Freese}.


Distinct differences in the $s-r$ plane have been realized for the
cases with different $n$ and $\beta$ for the modified polytropic
Cardassian universe. For \textbf{Case 1} and \textbf{Case 4}, the
crossing with LCDM happens at some $z_{\rm C}$, while the same state
never occurs for \textbf{Case 2} and \textbf{Case 5}. And
\textbf{Case 3} and \textbf{Case 6} just are two critical cases.
Also, the beginning points and the ending points are tightly related
to the shapes of the evolutionary trajectories. They seriously
depend on $n$ and $\beta$, especially $n$.
As $n$ and $\beta$ are found to be sensitive to the modified
polytropic Cardassian models, constraining the two parameters
exactly becomes a valuable task. We use the observational $H(z)$
data, the CMB data and the BAO data to make a combinational
constraint. We find that \textbf{Case 2} can be excluded at the
68.3\% confidence level and all the six cases are consistent with
the observational data at the 95.4\% confidence level. Recent other
constraints suggest some results far from consistent with each
other. For example, in the work of Wang et al., the choices of
$n=0.2$, $\beta =1$; $n=0.2$, $\beta =2$ and $n=0.2$, $\beta =3$ are
all consistent with SN Ia and CMB observations \cite{Wang}. The
first case corresponds to the quiessence with the equation of state
$\omega_{\rm Q}=-0.8$, and this case is consistent with \textbf{Case
5}. The latter two cases with the crossing at $z_{\rm C}$=0.29 and
0.26 respectively are in agreement with \textbf{Case 4}. And the
current values of the diagnostics are $s_{\rm 0}$=0.13, $r_{\rm
0}$=0.58 and $s_{\rm 0}$=0.16, $r_{\rm 0}$=0.45 for $n=0.2$,
$\beta$=2 and $n=0.2$, $\beta$=3 respectively. Nesseris \&
Perivolaropoulos \cite{Nesseris} suggested another result
$n=-23_{-9}^{+8}$, $\beta =0.025_{-0.010}^{+0.008}$ with a supernova
data set consisting of 194 SN Ia \cite{Tonry,Barris}. Both
\textbf{Case 1} and \textbf{Case 2} may be included within the given
error-bar. The best fitting values lie in \textbf{Case 1} and the
crossing with LCDM happens at $z_{\rm C}$=0.37. Meanwhile, the
present quantities are $s_{\rm 0}$=-0.29 and $r_{\rm 0}$=1.94. Evans
et al. provided a best fitting result $n=-0.94$ and $\beta =0.06$
with a simulated data set \cite{Evans}. This result is in agreement
with \textbf{Case 1}. The crossing happens in the far future, i.e.,
at $z_{\rm C}$=-0.999. And the current values $s_{\rm 0}$=0.76 and
$r_{\rm 0}$=0.54 indicate a large distance from LCDM. As constraints
still remain weak,
we expect for data with higher precision to provide more consistent
fitting results in future. We also hope theses statefinder
parameters can be determined more exactly and shed light on the
nature of the cosmological models.

\section{Acknowledgments}
We are very grateful to the anonymous referee for his valuable
comments that greatly improve this paper. And we are also grateful
to Xin Zhang for his helpful suggestions. Z. L. Yi would like to
thank Qiang Yuan and Jie Zhou for their kind help. This work was
supported by the National Science Foundation of China (Grants
No.10473002 and 10273003), the Scientific Research Foundation for
the Returned Overseas Chinese Scholars.

\end{document}